\documentstyle{article}
\begin{document}
\title{\Large\bf  Composite Polarons in Ferromagnetic Narrow-Band Metallic
Manganese Oxides \\}
\author{\large Liang-Jian Zou  \\
{\it  Institute of Solid State Physics, Academia Sinica,
       P.O.Box 1129, Hefei 230031, China}  \\
{\it  and CCAST (World Laboratory), P. O. BOX 8730, Beijing, 100083, China}\\
       H. Q. Lin \\
{\it  Department of Physics, the Chinese University of Hong Kong, Shatin, N.
       T. Hong Kong } \\
       and Q.-Q. Zheng\\
{\it  Institute of Solid State Physics, Academia Sinica,
        P.O.Box 1129, Hefei 230031, China}  \\
{\it and State Key Lab of Magnetism,
       Institute of Physics, Academia Sinica, Beijing, China }  \\ }
\date{ }
\maketitle
\large
\begin{center}
{\bf Abstract}
\end{center}
Moving electrons accompanied by Jahn-Teller phonon and spin-wave clouds may
form composite polarons in ferromagnetic narrow-band manganites.
The ground-state and the finite-temperature properties of such composite
polarons are studied in the present paper.
By using the variational method, it is shown that the energy of the system at
zero temperature decreases with the formation of the composite polaron and the
composite polaron behaves as a Jahn-Teller phononic polaron;
the energy spectrum of the composite polaron at finite temperatures is found 
to be strongly renormalized by the temperature and the magnetic field.
It is suggested that the composite polaron contributes significantly to
the transport and the thermodynamic properties
in ferromagnetic narrow-band metallic manganese oxides.

\vspace{1cm}

\newpage
\noindent {\bf I. INTRODUCTION} \\
Because of the electron-lattice interaction, an electron moving in a
polarized or a dynamically
distorted lattice surrounded by phonon clouds at low temperature
is damped and forms a phononic polaron [1]. Similarly, a moving electron
in ferromagnetic (or antiferromagnetic) background disturbs the nearby
local spin field and excites spin waves, therefore an electron surrounded
by spin-wave clouds can form a magnetic polaron [2]. One could anticipate
that in the presence of strong electron-lattice interaction and
electron-spin coupling, a moving electron clouded both by
phonons and spin waves can form a kind of new quasiparticle, which
we call a composite polaron. In the presence of a magnetic field and
for certain temperature, the physical
properties of a system with composite polarons differ from those of systems 
with either magnetic polarons or the phononic polarons. Since there exist
strong electron-phonon interactions, arising from the Jahn-Teller effect of
$MnO_{6}$, and strong electron-local spin interactions, arising from 
Hund's rule coupling, this kind of new
quasiparticle might be found in ferromagnetic manganese oxide materials,
such as the doped lanthanum manganites and neodymium manganites.

Recently the colossal magnetoresistance (CMR) effect has been found in
manganites, such as La$_{1-x}$Ca$_{x}$MnO$_{3}$ and
Nd$_{1-x}$Sr$_{x}$MnO$_{3}$ [3-6].
The huge magnetoresistance change of more than 3-6 orders in magnitude
has potential technical applications.
Other unusual transport, magnetic, and thermodynamic properties
and the microscopic mechanism of the CMR effect
have attracted great interest theoretically and experimentally.
The crystal and magnetic structures of La$_{1-x}$R$_{x}$MnO$_{3}$ systems
had been studied in 1950s by the x-ray crystallography and neutron diffraction
technique [7,8]. In manganites, an Mn ion is surrounded by
six O$^{2-}$ ions and forms an octahedron.
Due to the crystalline field effect, the 3d energy level of the Mn ion
is split into a low-lying triplet (t$_{2g}$) and a high-energy doublet
(e$_{g}$). Therefore, in La$_{1-x}$R$_{x}$MnO$_{3}$, the three d electrons
of Mn$^{3+}$ and Mn$^{4+}$ ions fill the d- t$_{2g}$ band (the filled band)
and form a localized core spin S=$\frac{3}{2}$ via strong intraatomic
Hund's rule coupling. The extra
d electrons in Mn$^{3+}$ fill the higher d-e$_{g}$ band and interact
with t$_{2g}$ electrons through strong Hund's coupling. These two bands are
separated by about 1.5 $eV$ [9]. The localized spins tend to align parallel
through the double exchange interaction between Mn$^{3+}$ and Mn$^{4+}$ ions
and form a ferromagnetic background [10, 11]. Electrons in the d-e$_{g}$
band
hop between Mn ions as itinerant ones and are responsible for the electric 
conduction in these systems. The degenerate d-e$_{g}$ doublet of Mn$^{3+}$
in MnO$_{6}$ octahedron will be further split due to the Jahn-Teller
effect [12, 13], the distortion of oxygen atoms reduces the symmetry of the
MnO$_{6}$ octahedron and stabilizes it. Electron moving together
with the distorted crystalline field (dynamical Jahn-Teller distortion)
will form a polaron. Recent experiments and theories [13-18] have confirmed 
the existence and the importance of the
Jahn-Teller distortion in La$_{1-x}$R$_{x}$MnO$_{3}$ compounds. Millis et al.
[13] suggested that the dynamic Jahn-Teller effect plays a crucial role for the
physical properties of manganites. In the meantime, doped
$ La_{1-x}R_{x}MnO_{3}$ $(0.1<x<0.5)$  are ferromagnetically ordered at
temperatures below the Curie temperature T$_{c}$. The exchange coupling between
the mobile electron and ordered localized spins can form a
magnetic polaron. Therefore the motion of an electron at the e$_{g}$ level
in lanthanum manganites may form a composite quasiparticle, with the
coexistence of phononic and magnetic polaron simultaneously in the same place,
this called the composite polaron.

The composite polaron (CP) in the present situation is small in the
metallic regime.
An earlier study [19] had shown that the lattice deformation
can stabilize small magnetic polaron, so the CP  formed in ferromagnetic
narrow-band metallic manganese-oxides is stable. In this paper
in Sec. II and Sec. III, the framework of the composite polaron is developed
for perovskite-type manganese-oxides and its dependence on the temperature
and the magnetic field are discussed.
We found that the formation of a CP in manganese oxide systems is favorable
in energy, and propose that the CP plays an essential role for the
transport and the thermodynamic properties of lanthanum manganites.
The formation of the CP could be responsible for the microscopic mechanism
of the CMR effect. Finally, we draw conclusions in Sec. IV. \\

\noindent {\bf II. Composite Polaron at Zero Temperature}

   We first discuss the possibility of the formation of CP in ferromagnetic
narrow-band manganese-oxides at zero temperature.
The typical physical processes in manganites can be decomposed into three
parts: the electron-electron interaction (the hopping process and the
Coulomb interaction), H$_{0}$, the
electron-lattice interaction (Jahn-Teller effect), H$_{e-ph}$,
and the interaction between localized spins and mobile electrons (Hund's
coupling), H$_{e-m}$. So the Hamiltonian is:
\begin{equation}
       H=H_{0}+H_{e-m}+H_{e-ph}
\end{equation}
\begin{equation}
 H_{0}=\sum_{i a \sigma} [\epsilon^{a}_{d}-
 \sigma \mu_{B}B] d^{\dag}_{i a\sigma}d_{ia \sigma}+
\sum_{<ij> ab \sigma} t^{ab} d^{\dag}_{i a\sigma}d_{jb \sigma}
\end{equation}
\begin{equation}
   H_{e-ph}=  \sqrt{\alpha \hbar\omega} \sum_{ia}
             n_{ia}\beta_{a} [b^{\dag}_{i}+b_{i}]+\sum_{i}
            \hbar \omega b^{\dag}_{i}b_{i}
\end{equation}
\begin{equation}
   H_{e-m} = -J_{H} \sum_{ia\mu \nu}
       {\bf S}_{i} \cdot d^{\dag}_{ia\mu} {\bf \sigma}_{\mu \nu} d_{ia\nu}
  - \sum_{<ij>} {\it A}_{ij} {\bf S}_{i} \cdot {\bf S}_{j}-
       \sum_{i}g\mu_{B}BS^{z}_{i}
\end{equation}
In Eq.(2), d$^{\dag}_{ia\sigma}$ creates an e$_{g}$ electron at site
${\bf R}_{i}$ at level $a$ with spin $\sigma$, $t^{ab}$ denotes the hopping
integral of the e$_{g}$ electron from one site to its nearest-neighbor,
$\epsilon_{d}$ is the site energy of the mobile electron with respect to
the chemical potential, $a$ (or $b$) denotes the energy-level index
of Jahn-Teller splitting in e$_{g}$ level. In Eq.(3), b$^{\dag}_{i}$ creates
a Jahn-Teller phonon at site ${\bf R}_{i}$ with mode $\hbar \omega$,
$\alpha$ represents the electron-phonon coupling constant,
$\beta_{a}$ is a two-component constant vector, it is -1 or 1 for the low or
the high level of the Jahn-Teller splittings, respectively.
Several experiments have shown that in the metallic regime, the magnetic
ordering of the local spins is of Heisenberg ferromagnetic type [20, 21].
In Eq.(4), ${\it A}_{ij}$ is the effective ferromagnetic
exchange constant between manganese spins with only the nearest-neighbor
interaction being considered, -g$\mu_{B}$B represents the Zeemann energy in
magnetic field {\bf B}. The mobile electron is scattered from state
$i\nu$ to state $i\mu$ by the localized spin ${\bf S_{i}}$ due to the Hund's
coupling J$_{H}$ between the mobile electrons and the core spins.
As suggested by Kubo and Ohata [22], Millis et al [13] and Zang et al [14],
J$_{H} >> t^{ab}$, 
so the bandwidth of the metallic manganites is narrow.

In doped perovskite manganites, the double exchange interaction due to
the hole hopping and the strong Jahn-Teller electron-phonon coupling is
limited to a range of a few lattice constant. This fact suggests that
small polaron picture is suitable for the CMR materials.
Through a canonical transformation, H$^{'}=e^{-S}$H$e^{S}$, one could
eliminate the linear term in phonon degrees of freedom in Eq. (1). By chosing
S=$\sum_{i} \sqrt{\alpha / \hbar \omega}$ n$_{i\alpha} \beta_{\alpha}$
($b^{\dag}_{i}- b_{i}$), an effectively attractive electron-electron
interaction is introduced and the Hamiltonian Eq. (1) can be rewritten as:
\begin{eqnarray}
H^{'} &=& \sum_{ia\sigma} \epsilon^{a}_{i\sigma}
                          d^{\dag}_{ia\sigma} d_{ia\sigma}
        + \sum_{<ij>ab\sigma}t^{ab} d^{\dag}_{ia\sigma} d_{jb\sigma}
                          \hat{X}^{\dag}_{ia}\hat{X}_{jb} \nonumber\\
      &-& \sum_{ia,b} \alpha n_{ia}\beta_{a}n_{ib}\beta_{b}
 +\sum_{i} \hbar\omega_{i} b^{\dag}_{i}b_{i} +H_{e-m} ~.
\end{eqnarray}
Where $\epsilon^{a}_{i\sigma}= \epsilon^{a}_{d}-\mu-\sigma \mu_{B}B$,
and $\hat{X}^{\dag}_{ia}$ =
exp$[\sqrt{\alpha/\hbar\omega} \beta_{a}(b_{i}-b^{\dag}_{i})]$.
In the present studies we are only interested in the ground state of the CP
so it is reasonable to assume that the mobile electrons stay in the lower level
of Jahn-Teller splittings and we will drop the index $a$ and $b$ in Eq. (5).
The term \( -\sum_{ia,b} \alpha n_{ia}\beta_{a}n_{ib}\beta_{b} \) is reduced
to a renormalization shift of the bare energy level by $-\alpha$ and
the energy shift of the electrons from the electron-phonon interaction,
$\alpha$, is absorbed into $\epsilon^{a}_{d}$.

   We choose a set of basis function to construct the variational ground-state
wavefunction of the CP at T=0 K:
\begin{equation}
   |\psi_{i}(S^{z}_{0})> =  \sqrt{\frac{S+S^{z}_{0i}+1}{2S+1}}
[d^{\dag}_{i\uparrow}+\frac{d^{\dag}_{i\downarrow}S^{\dag}_{i}}{S+S^{z}_{0i}+1}]
\Pi_{j=1}^{N} |S_{j}S^{z}_{j} > |0>_{e-ph}
\end{equation}
where $\Pi_{j=1}^{N} |S_{j}S^{z}_{j} >$ denotes the ferromagnetic background
and $|0>_{e-ph}$ represents the electron and the phonon vacuum state.
\( S^{z}_{0i} \) denotes the mean value of z-component of the local spin at
R$_{i}$. This basis set is constructed from the wavefunction of the vacuum
state of phonon, $ |0>_{ph} $, and that of the magnetic polaron (Ref. [2]):
$|\chi>$=$\sqrt{\frac{S+S^{z}_{0i}}{2S+1}}$ [d$^{\dag}_{i\downarrow}$
- d$^{\dag}_{i\uparrow}S^{-}_{i}/(S+S^{z}_{0i})]$
$\Pi_{j=1}^{N} |J_{j}J^{z}_{j} >$ $ |0>_{e}$.
If we let $\alpha \rightarrow 0$ or J$_{H} \rightarrow 0 $ in (5), the
problem reduces to that of the magnetic polaron or
the phononic polaron, respectively.

The ground state wavefunction of the CP can be constructed in terms of the
basis set, $|\psi_{i}(S^{z}_{0})>$, through a linear combination:
\begin{equation}
   |G> =\sum_{iS^{z}_{0}} c_{i}(S^{z}_{0}) |\psi_{i}(S^{z}_{0})>
\end{equation}
here $c_{i}(S^{z}_{0})$ is the variational coefficient of the
ground-state wavefunction, \( S^{z}_{0} \) represents all spin variables.
Acting the Hamiltonian H$^{'}$ on $|G>$ gives
to the ground-state energy of the CP:  E$_{g}$$<G|G> =<G|H^{'}|G>$.
Minimizing the above expression with respect to the coefficients
$c_{i}(S^{z}_{0}$), one gets:
\begin{equation}
     E_{g} c_{i}=-J_{H}Sc_{i}+te^{-\frac{p}{2}}
\sum_{\delta}\frac{\sqrt{(S+S^{z}_{0i}+1)
(S+S^{z}_{0i+\delta}+1)}}{2S+1}c_{i+\delta}
\end{equation}
\[ + \frac{S+ S^{z}_{0i+\delta}+1}{2S+1}[
\epsilon_{\uparrow}+\epsilon_{\downarrow}\frac{S-S^{z}_{0i}}{S+S^{z}_{0i
+\delta}+1}]c_{i}+\frac{S-S^{z}_{0i}}{2S+1}[-g\mu_{B}B-2A\sum_{\delta}
S^{z}_{0i+\delta}]  ~.\]
After taking Fourier transformation, the nontrivial solution of the
coefficients c$_{i}(S^{z}_{i})$ gives rise to the ground-state energy spectrum
of the CP:
\begin{equation}
  E_{g}(k) =\epsilon_{d}-J_{H}S+zte^{-\frac{p}{2}}\gamma(k)
\frac{\sqrt{(S+S^{z}+1)(S+S^{z}_{\delta}+1)}}{2S+1}
\end{equation}
\[ -\frac{2S}{2S+1}[\mu_{B}B+\frac{g}{2S}\mu_{B}B+A\sum_{\delta}
S^{z}_{\delta}]  \]
where $p=\alpha/\hbar\omega$ is the relative strength of the
electron-phonon interaction, z is the partition number, and $\gamma(k)$
the structure factor.
The physical meaning of each term in the above expression is obvious:
the first two terms, $\epsilon_{d}-J_{H}S$, came from the intraatomic
interaction, and the third term describes the renormalized energy dispersion.
The electron-phonon interaction and the spin-electron coupling narrow the bare
energy spectrum of the electrons, zt$\gamma(k)$, by a factor of exp$(-p/2)$ and
a factor of \( \frac{\sqrt{(S+S^{z}_{0}+1)(S+S^{z}_{0\delta}+1)}}{2S+1} \),
respectively.
However one notices that at the absolute zero temperature, the system is
ferromagnetically ordered and there is no spin excitation, S$^{z}$=S,
so that $\frac{\sqrt{(S+S^{z}_{0i}+1)(S+S^{z}_{0i+\delta}+1)}}{2S+1}=1$.
Therefore the CP behaves as phononic polaron at zero temperature.
This point lies in the following fact that at zero temperature, there
does not exist any spin deviation in ferromagnets so the electron is only
accompanied by dynamical distorted lattice field.
However the zero-temperature CP differs from the Jahn-Teller polaron in the
intraatomic Hund's coupling (the second term in Eq. (9)) and the effective
Zeemann energy of the CP (the last term in Eq. (9)).
It can be seen that the formation of the CP lowers the energy of the system
significantly, so it is more favorable in energy.

The presence of strong Jahn-Teller electron-phonon coupling in ferromagnetic 
perovskite manganites has been confirmed by several experiments [15-18].
By using neutron powder-diffraction data [15], it is shown that the static 
distortion of oxygen around manganese resulting from Jahn-Teller 
electron-phonon coupling is about 0.12 $\AA$ in
La$_{1-x}$Ca$_{x}$MnO$_{3}$ (x $\approx$ 0.2).
The direct support of the
presence of electron-phonon coupling came from the oxygen isotope
experiments done by Zhao et al. [23].
It is believed that the ferromagnetic coupling between manganese spins comes
from the double-exchange interaction, mediated through oxygen atoms,
so the motion of oxygen atoms will affect the double-exchange strength,
hence the ferromagnetic coupling, by a factor e$^{-p(1/2+<n_{ph}>)}$
(here $<n_{ph}>$ is the mean phonon number at finite temperatures, see 
next section).
Since p $\sim 1/\omega$ $\sim$ $M^{1/2}$, M is the mass of 
oxygen, the heavier the
oxygen nucleus is, the weaker the coupling is. Therefore, the presence of
a strong electron-phonon interaction will lead to a significant decrease
of the Curie temperature. Zhao et al. showed that the oxygen
isotope exponent, $\alpha_{o}$=-d$lnT_{c}$/d$lnM_{o}$, is as high as 0.85 for
La$_{0.8}$Ca$_{0.2}$MnO$_{3}$, suggesting the existence of strong
electron-phonon interaction. Accordingly, one could
estimate that the relative strength of electron-phonon interaction, p,
is about $1 -2$, which belongs to the strong coupling regime.

On the other hand, besides the strong electron-phonon 
interaction, the spin-electron interaction is also pretty strong.
In the model Hamiltonian Eq.(4), the Hund's coupling between the carrier and
the local spin is about 5 eV, much larger than the conduction bandwidth 2zt
($\approx$ 2.0 eV [13]). Experiments done by Kuster [3], von 
Helmolt [4], Jin [5] and other groups
proposed that spin polaron may dominate the electric transport.
Such viewpoint is
supported by the fact that the logarithm of the conductivity of
La$_{1-x}$Ca$_{x}$MnO$_{3}$ exhibits T$^{-1/4}$ dependence on temperature
[24], which suggests a typical spin polaron
character in the transport properties of La$_{1-x}$Ca$_{x}$MnO$_{3}$.
\\

\noindent {\bf III. Properties of Composite Polaron at Finite Temperature}

  With the raise of temperature, more and more phonons and spin-waves are
excited, whereas the increase of an external magnetic field decreases
the spin-wave excitations.
Thus
the formation and the properties of the CP can be heavily affected by
the external magnetic field {\bf B} and temperature T.

   At finite temperatures, the thermal excitation of phonon and spin waves
become more and more important, their occupations obey the Bose-Einstein
distribution law.
We choose the following basis function for the CP at temperature T:
\begin{equation}
   |\phi_{i}(S^{z}_{0})> =  \sqrt{\frac{S+S^{z}_{0i}}{2S+1}}
[d^{\dag}_{i\uparrow}+\frac{d^{\dag}_{i\downarrow}S^{\dag}_{i}}{S+S^{z}_{0i}}]
|0>_{e} |n_{i}>_{ph} |m_{i}>_{m}
\end{equation}
where $|0>_{e}$ denotes the electron vacuum state, $|n_{i}>_{ph}$ the 
thermal-equilibrium phonon
state and $|m_{i}>_{m} $ the magnetic excitation state at site R$_{i}$ at
temperature T under magnetic field {\bf B}. Denoting $a^{+}_{i}$ (or $a_{i}$)
as the creation (or the annihilation) operator of the CP, then
the wavefunction of the CP can be expressed as:
\begin{equation}
   |\Phi_{i}> =\sum_{i} a_{i} |\phi_{i}(S^{z}_{0})> ~.
\end{equation}
Expressing the Hamiltonian (5) in terms of the CP operators, we obtain
\begin{equation}
   \tilde{H} =\sum_{i,j} a^{\dag}_{j} a_{i}
<\phi_{j}(S^{z}_{0})|H^{'}|\phi_{i}(S^{z}_{0})>
\end{equation}
After a tedious calculation with the use of Fourier transformation, one gets:
\begin{eqnarray}
   \tilde{H} &=& \sum_{k} [\epsilon_{d}-J_{H}S
                 -\frac{2S}{2S+1}(\mu_{B}B+\frac{g}{2S}\mu_{B}B+A\sum_{\delta}
                 S^{z}_{\delta})] a^{\dag}_{k} a_{k} \nonumber\\
             &+& \sum_{k} zte^{-p(<n_{B}>+\frac{1}{2})}[1+
                 \frac{1}{2S+1}\sum_{q}(\frac{2S}{2S+1}\gamma_{q} -1)<m_{q}>]
                 \gamma(k) a^{\dag}_{k} a_{k}
\end{eqnarray}
where $<n_{B}>=1/[exp(\hbar\omega/k_{B}T)-1]$ and $<m_{q}>$=
$1/[exp(\hbar\Omega_{q}/k_{B}T)-1]$ denote the thermal-equilibrium mean
occupations of the phonon and of the spin waves at temperature T, 
respectively; here $\hbar\Omega_{q}$ is the energy spectrum of the spin 
waves in double-exchange ferromagnets,
$\hbar\Omega_{q}=g\mu_{B}B+2zA(1-\gamma_{q})$. 
In the calculation, the linear approximation of the Holstein-Primakoff
transformation:
\[ S^{\dag}=\sqrt{2S}c,~~ S^{-}=\sqrt{2S}c^{\dag},~~S^{z}=S-c^{\dag}c   \]
was adopted. The average energies of the free phonons and of the free spin
waves were not included.

Compare the dispersion of the CP in Eq. (13) with the bare energy spectrum 
$zt\gamma(k)$ of electrons, the bandwidth of the carriers at finite temperature
T is renormalized by a factor of Z,
\begin{equation}
 Z = e^{-p(<n_{B}>+\frac{1}{2})}[1-\frac{1}{2S+1}\sum_{q}(1-\frac{2S}{2S+1}
\gamma(q)) <m_{q}>] ~.
\end{equation}
One finds that the renormalized factor Z
consists of two parts: the phonon renormalized part Z$_{ph}$ and the spin-wave
part Z$_{m}$, Z=Z$_{ph}$Z$_{m}$.
Z exhibits strong temperature- and magnetic field-dependence.
The phononic part,
\[ Z_{ph}=exp[-p (<n_{B}>+\frac{1}{2})]     ~, \]
depends only on the temperature. The increase of the electron-phonon 
interaction and temperature will narrow the conduction band further.
The magnetic part,
\[  Z_{m}= 1-\frac{1}{2S+1}\sum_{q}[1-\frac{2S}{2S+1} \gamma(q)] <m_{q}>  ~, \]
depends both on the temperature and the magnetic field.
With the increase of temperature, the occupation of spin waves, $<m_{q}>$,
becomes large. Since $(1-\frac{2S}{2S+1} \gamma(q))$ is always positive, so
the factor Z$_{m}$ becomes small.
Therefore in the region where temperature is below T$_{c}$, the bandwidth of
the CP becomes narrower and narrower with the increase of temperature and
the CP may be more easily trapped, so the mobility of the CP becomes smaller.
Under strong magnetic field, the increase
of field strength depresses the excitation of spin waves, reduces the 
number of spin waves surrounding the electron, thus the bandwidth of
the CP becomes broader with external magnetic field and the mobility of
the CP becomes larger.
These properties coincide with experimental observations for resistivity.
When T approaches the Curie temperature from below, the number of spin wave
excitations reach its maximum and the factor Z$_{m}$ approaches its minimum.
When T $> T_{c}$, the above treatment of the linear spin wave approximation
(via the Holstein-Primakoff transformation) is no longer true since
the long-range ferromagnetic order does not exist anymore and
one has to deal with the spin operators rather than the spin wave operators.
Recent experimental observations [25] showed that the collective
spin excitations also exist above the Curie temperature which suggested that
the composite polaron can preserve for temperatures above T$_{c}$.

   An interesting property of the CP is its transport properties in
the presence of magnetic field B at finite temperature T,
which is closely related to the transport properties of
lanthanum manganite, especially the CMR effect.
It is found that the magnetoresistance behavior of lanthanum manganites
can be understood qualitatively in the present theory of composite polaron.
The electric resistivity can be expressed as
\[     \rho =1/(ne\mu), ~~~~~ ~\mu =e \tau/m^{*}    \]
where $\mu $ is the mobility of the CP, $m^{*}$ is the effective mass 
which is proportional to the inverse of the bandwidth.
According to Eq.(13) and (14), one can determine the mobility through
the effective mass,
$\mu \propto 1/m^{*} \propto Z_{m}Z_{ph}$.
At temperatures below the Curie temperature T$_{c}$, there exist both
the spin-wave and the Jahn-Teller phonon excitations.
As temperature arises, more and more spin waves and phonons are excited
and the mobility of the CP becomes smaller. Consequently, the resistivity
increases. The electric conductivity decreases with the increase of
temperature and reaches its maximum near the Curie point.
Above T$_{c}$, the long-range magnetic ordering disappears
and only the Jahn-Teller phononic polaron plays a role.
Like usual polarons, the hopping probability of the polaron to the nearby
sites increases with the lift of temperature,
so the resistance declines with the increase of T in high temperature region.
Due to strong polaron self-trapping effect and its thermal activation,
one would expect that a great change of the resistance with the increase of
temperature, and that the depression of the magnetic field to the local spin
fluctuation will reduce the bonding strength of the composite polaron, hence
decrease the resistivity. From the preceding discussions, it is found that
the narrowing of the electronic bandwidth is a fundamental
process controlling the transport properties of conduction electrons in
ferromagnetic narrow-band metallic manganites,
a point supported by recent experiment [26].

  One could estimate the magnetoresistance change
through the mobility of the CP. The mobility variation comes
from two parts, the effective mass and the scattering lifetime:
\begin{equation}
      \frac{\mu}{\mu_{0}} = \frac{m}{m^{*}} \frac{\tau}{\tau_{0}}
\end{equation}
here $\mu_{0}$ and $\tau_{0}$ denote the mobility and the scattering 
lifetime of free carrier.
For typical electron-phonon coupling, p$ \approx 1$, and
the ferromagnetic coupling, 4zA =30 $meV$ [20] when temperature ranges
from 50 to 250 K, the variation
of the contribution of the effective mass to the mobility of the composite
polaron decreases about one order in magnitude, or, m/m$^{*}$ $\sim 0.1$. The 
effect of magnetic field is not so significant as temperature. At T=250 K, the
the mobility contribution from the effective mass increases by two times
when the magnetic field increases from
zero to 8 Tesla. The reason is that the magnetic field only affects
Z$_{m}$, however, temperature affects both Z$_{ph}$ and Z$_{m}$.
In fact, the huge change of the resistivity of the CP with
magnetic field and temperature may come from the variation of the
scattering lifetime, which can change several orders in magnitude and
drive a metal-insulator-like transition in doped manganites. 
Detail studies is tedious and will be published elsewhere.\\

\noindent {\bf IV. Conclusion}

   To summarize, a theory of the composite polaron is developed. It is suggested
that the composite polaron can be realized in lanthanum manganites, and
contribute significantly to the unusual ground state, the thermodynamic,
and the transport properties of manganites.
\\

{\bf Acknowledgements:} Liang-Jian Zou thanks  Li
Zheng-Zhong for discussions, Yu Lu and the invitation of ICTP in Trieste,
Italy, for hospitality. This work is supported by the
Grant of the NNSF of China and the Grant of CAS,
and by the Direct Grant for Research from the Research Grants Council (RGC)
of the Hong Kong Government.

\newpage
\begin{center}
REFERENCES
\end{center}
\begin{enumerate}
\item S. I. Peker, {\it Zh. Esksp. Teor. Fiz.}, {\bf 16}, 341 (1946)
\item E. L. Nagaev, {\it Sov. Phys}. JETP, {\bf 29}, 545 (1969); {\bf 31}, 682
      (1969).
\item R. M. Kusters, J. Singleton, D. A. Keen, R. McGreevy and W. Hayes,
       {\it Physica} {\bf B155} 362 (1989).
\item R. Von Helmolt, J. Wecker, B. Holzapfeil, L. Schultz and K. Samwer,
       {\it Phys. Rev. Lett.} {\bf 71} 2331 (1993);
       {\it J. Appl. Phys.} {\bf 76}, 6925 (1994).
\item S. Jin, T. H. Tiefel, M. McCormack, R. A. Fastnacht, R. Ramesh, and
      L, H, Chen  {\it Science}. {\bf 264} 413 (1994).
\item G. C. Xiong, Q. Li, H. L. Ju, S. N. Mao, L. Senpati, X. X. Xi, R. L.
      Greene, and T. Venkatesan, {\it Appl. Phys. Lett} {\bf 66}, 1427 (1995).
\item R. C. Vickery and A. Klann, {\it J. Chem. Phys.}, {\bf 27}, 1161 (1957).
\item  Goodenough, {\it Phys. Rev.} {\bf 29}, 111 (1957)
\item J. M. D. Coey, M. Viret and L. Ranno, {\it Phys. Rev. Lett.},
      {\bf 75}, 3910 (1995).
\item C. Zener, {\it Phys. Rev.}, {\bf 81}, 440 (1951); {\bf 82}, 403
      (1951).
\item P. G. De Gennes, {\it Phys. Rev.}, {\bf 100}, 564 (1955); {\bf 118}, 141
      (1960).
\item H. A. Jahn, {\it Proc. Roy. Soc.}, {\bf 164},117 (1938)
      Y. E. Perlin and M. Wagner, The Dynamical Jahn-Teller Effect in Localized
      Systems, North-Holland (1984)
\item A. J. Millis , P. B. Littlewood, and B. I. Shraiman, {\it Phys.
      Rev. Lett} {\bf 74}, 3407 (1995).
\item Jun Zang, A. R. Bishop and H. Roder, {\it Phys. Rev.},
      {\bf B53}, R8840 (1996).
\item S. J. L. Billinge, R. G. DiFrancesco, G. H. Kwei, J. J. Neumeier, J. D.
      Thompson, {\it Phys. Rev. Lett.} {\bf 77} 715 (1996).
\item M. C. Martin, G. Shirane, Y. Endoh and K. Hirota, Y. Moritomo and Tokura
      {\it Phys. Rev.}, {\bf B53}, 14285 (1996).
\item T. A. Tyson, J. Mustre de Leon, S. D. Conradson, A. R. Bishop, J. J.
        Neumeier, H. Roder and Jun Zang.
       {\it Phys. Rev.}, {\bf B53}, 13985 (1996).
\item I. Solovyev, N. Hamada and K. Terakura,
      {\it Phys. Rev. Lett.} {\bf 76}, 4825 (1996).
\item M. Umehara, {\it Phys. Rev}, {\bf B27}, 5669 (1983)
\item J. W. Lynn, R. W. Erwin, J. A. Borchers, Q. Huang, A. Santoro, J-L. Peng
      and Z. Y. Li, {\it Phys. Rev. Lett.}, {\bf 76}, 4046 (1996).
\item T. G. Perring, G. Aeppli, S. M. Hayden, S. A. Carter, J. P.
Remeiza, and S.-W. Cheong, {\it Phys. Rev. Lett.}, {\bf 76}, 4046 (1996).
\item K. Kubo and N. Ohata,  {\it J. Phys. Soc. Jpn.}, {\bf 33}, 21 (1972).
\item Guo-meng Zhao, K. Conder, H. Keller, K. A. Muller. {\it Nature}, 
      {\bf 381}, 676 (1996).
\item J. M. De Terasa, M. R. Ibarra, J. Blasco, J. Garcia, C. Marquina, 
      P.A. Algarabel, Z. Arnold, K. Kamenev, C. Ritter, and R. Von Helmolt, 
      {\it Phys. Rev.}, {\bf 54}, 1187 (1996).
\item S. B. Oseroff, M. Torikachvili, J. Singley, S. Ali, S.-W. Cheong and S.
       Schultz, {\it Phys. Rev}, {\bf B53}, 6521 (1996)
\item J. Fontcuberta, B. Martinez, A. Seffar, S. Pinol, J. L. Garciamunoz,
      X. Obrasors,  {\it Phys. Rev. Lett.}, {\bf 76}, 1122 (1996).

\end{enumerate}

\end{document}